\documentclass[letterpaper, 10 pt, conference]{ieeeconf}

\IEEEoverridecommandlockouts                              

\usepackage{graphicx} 
\usepackage{amsmath,bm,times} 
\usepackage{amssymb}  
\usepackage{bbm}
\usepackage{bm}
\usepackage{tikz}
\usetikzlibrary{shapes,arrows,backgrounds,fit,positioning}
\usepackage{subfigure}
\usepackage{cite}
\usepackage{booktabs} 
\usepackage[linesnumbered, ruled, vlined]{algorithm2e}
\usepackage{tabularx}
\usepackage{csquotes}
\usepackage{balance}

\newtheorem{remark}{Remark}
\newtheorem{theorem}{Theorem}

\newtheorem{corollary}{Corollary}

\newtheorem{example}{Example}

\newcommand{\rev}[1]{{\color{black}#1}}

\title{\Large \bf Tuning Cooperative Behavior in Games with Nonlinear Opinion Dynamics
\thanks{Supported by funding from King Abdullah University of Science and Technology (KAUST), ONR grant N00014-19-1-2556, ARO grant W911NF-18-1-0325, DGAPA-UNAM PAPIIT grant IN102420, Conacyt grant A1-S-10610,  and  
NSF Graduate Research Fellowship  
DGE-2039656.}}
\author{Shinkyu Park, Anastasia Bizyaeva, Mari Kawakatsu, Alessio Franci, and Naomi Ehrich Leonard

  \thanks{S. Park is with King Abdullah University of Science and Technology (KAUST), Computer, Electrical and Mathematical Science and Engineering 
  Division, Thuwal 23955-6900, Saudi Arabia. {shinkyu.park@kaust.edu.sa}}

  \thanks{A. Bizyaeva and N. E. Leonard are with the Department of Mechanical and Aerospace Engineering, and M. Kawakatsu is with the Program in Applied and Computational Mathematics, Princeton University, Princeton, NJ 08544, USA. \{bizyaeva, naomi, mari.kawakatsu\}@princeton.edu}

  \thanks{A. Franci is with Mathematics Department, National Autonomous University of Mexico, 04510 Mexico City, Mexico. {afranci@ciencias.unam.mx}}
}

\IEEEoverridecommandlockouts
\begin{document}

\maketitle

\begin{abstract}
\rev{We examine the tuning of cooperative behavior in repeated multi-agent games using an analytically tractable, continuous-time, nonlinear model of opinion dynamics. Each modeled agent updates 
its real-valued opinion about 
each  available strategy in response to  payoffs 
and other agent opinions, as observed over a network.} 
We show how the model provides a principled and systematic means to investigate behavior of agents that select strategies using rationality and reciprocity, 
key 
features of 
human decision-making in social dilemmas. 
 \rev{For two-strategy games, we use bifurcation analysis to prove conditions for the bistability of two equilibria and conditions for the first (second) equilibrium to reflect all agents favoring the first (second) strategy. 
We prove how model parameters, e.g., level of attention to opinions of others (reciprocity), network structure, and payoffs, influence dynamics and, notably, the size of the region of attraction to each stable equilibrium. 
We provide insights 
by examining the tuning of the bistability of mutual cooperation and mutual defection and their regions of attraction for the repeated prisoner's dilemma and the repeated multi-agent public goods game.} 
Our results generalize to games with
more strategies, heterogeneity, and additional feedback dynamics, such as those designed to elicit cooperation. 
\end{abstract}

\section{Introduction} \label{sec:intro}
\rev{
Sociologists, political scientists, and economists have long argued that \textit{reciprocity} is key to promoting cooperation \cite{10.2307/2092623,Axelrod84,RePEc:eee:eecrev:v:42:y:1998:i:3-5:p:845-859}. 
Computer simulations have shown that reciprocal strategies can elicit mutual cooperation in repeated games: the winning strategy for the repeated prisoner's dilemma in Axelrod's tournaments was Tit-for-Tat (TFT), where an agent reciprocates the opponent's strategy in the previous round; more generally, successful strategies were \textit{nice}, \textit{forgiving}, \textit{provocable}, and \textit{clear} \cite{Axelrod84}. 
Subsequent laboratory studies have revealed that humans in fact 
employ such reciprocity-based rules 
in repeated interactions 
\cite{
doi:10.1146/annurev.soc.24.1.183,doi:10.1098/rsos.182142, Mao2017}.
However, the observed reciprocity cannot be recapitulated by game-theoretic models of rational, payoff-maximizing agents, which, in contrast to the experiments, predict convergence toward mutual defection, i.e., the Nash equilibrium in a social dilemma.
}

\rev{
Here we investigate the tuning of cooperative behavior, including mutual cooperation or coordination, in repeated games among agents that rely on both rationality and reciprocity.}
\rev{Our first key contribution is a new framework for studying multi-agent repeated games using 
the \textit{nonlinear opinion dynamics model}} \cite{Bizyaeva2021} (see also \cite{8265165}) 
in which agents' strategic decisions 
depend not only on 
payoffs, as in rationality models \cite{9022871, 10.5555/3219358.3219364}, but also 
on \textit{social interactions} that enable agents to observe  strategy preferences (opinions) of other agents. 
We show how the social interaction term, 
formulated as a saturation function of observed opinions, 
provides a representation of reciprocity \rev{and a means to tune cooperation (or coordination) in social dilemmas.} 

{\color{black}
Our second key contribution leverages analytical tractability of the model: we prove conditions for bistability of two equilibria for repeated two-strategy games in which multiple agents observe the opinions of others over a fixed network. We also show conditions under which each equilibrium corresponds to all agents favoring one of the two strategies. 
Our proof relies on a bifurcation analysis that builds on the results of \cite{Bizyaeva2021}. We  prove how the  bistability of equilibria and the regions of attraction depend on level of attention to observed opinions (reciprocity), network structure, payoffs, and other model parameters. 
We apply our theory to the two-agent prisoner's dilemma and the multi-agent public goods game to present further insights on how mutual cooperation emerges through social interaction (reciprocity) and  how the predicted likelihood of cooperation can be tuned. Our results apply analogously to tuning coordination in games like the Stag Hunt.
Our analytical results complement the large literature on reciprocity-based decision-making   \cite{Axelrod84} 
that evaluates agents' long-term interaction with computer simulations.} 

{\color{black}
Most models of opinion dynamics in the literature use an opinion updating process that relies on a linear weighted average of exchanged opinions, as in the original work of DeGroot 
\cite{10.2307/2285509}. 
The nonlinear opinion dynamics model of \cite{Bizyaeva2021} instead applies a saturation function to exchanged opinions, making the updating process fundamentally nonlinear and thus allowing for multistability of equilbria, a key aspect of our project. For a comprehensive review of, and comparison with, other opinion dynamics models see \cite{Bizyaeva2021}. Our investigation of the means to tune cooperation in social dilemmas is also distinguished from works such as \cite{7024142, BAUSO2018204} that examine opinion dynamics using game-theoretic approaches.

Our approach is also distinguished from the investigations in \cite{Bizyaeva2021}: evolving opinions, which represent strategy preferences,  depend not only on saturated opinion exchange  but also on the payoff mechanism of the game. 
Our results are also new: they explain 
the emergence of mutual cooperation (or coordination) in social dilemmas as one of two bistable equilibria that arise through a pitchfork bifurcation. 
}


In \S\ref{sec:opinion_dynamics_in_games}, we introduce the nonlinear opinion dynamics model 
and  show  
how it recovers rationality and reciprocity. 
In \S\ref{sec:analysis}, \rev{for two-strategy games, we prove the bistability of equilibria
and expressions for the tunability of those equilibria and their corresponding regions of attraction in terms of system parameters. 
We apply the theory to the prisoner's dilemma and public goods game.}  
In \S\ref{sec:simulations} we use numerical simulations to illustrate the theoretical predictions on the tuning of cooperation.
In \S\ref{sec:conclusion}, we discuss extensions and generalizations. 

\section{Opinion Dynamics in Games} \label{sec:opinion_dynamics_in_games}
Consider an $N_a$-agent decision-making problem where each agent selects a strategy, \rev{continuously in time $t$,} from the set $\{1, \cdots, N_s\}$ of $N_s$ available strategies. 
Each agent performs a probabilistic choice of strategy where \rev{$x_i(t) \in \mathbb X_i$} is the probability distribution for the strategy selection at time $t$ of agent $i$ and $\mathbb X_i$ is the probability simplex in $\mathbb R^{N_s}$. The $j$-th element $x_{ij}$ of $x_i$ is the probability that agent~$i$ selects strategy~$j$.  Following convention 
in game theory \cite{Sandholm2010Population-Game}, $x_i$ is the \textit{mixed strategy} of agent~$i$ and $x = (x_1, \cdots, x_{N_a}) \in \mathbb X$ is the \textit{mixed strategy profile}, where $\mathbb X = \mathbb X_1 \times \cdots \times \mathbb X_{N_a}$.

The mixed strategy $x_i(t)$ is defined by the logit choice function \cite{10.5555/3219358.3219364} and depends on agent~$i$'s \textit{opinion state} \rev{at time $t$, $\bar z_i(t) = \left( \bar z_{i1}, \cdots, \bar z_{iN_s} \right)(t) \in \mathbb R^{N_s}$}, as follows:
\begin{align} \label{eq:logit_choice}
  x_{ij} = \sigma_j \left( \bar z_i \right) = \frac{\exp\left( \eta^{-1} \bar z_{ij} \right)}{\sum_{l=1}^{N_s} \exp\left( \eta^{-1} \bar z_{il} \right)},
\end{align}
where the positive constant $\eta$ is called the noise level \cite{HOFBAUER200747} or rationality parameter \cite{CHEN199732}.\footnote{For simplicity, we assume that $\eta$ is identical across the agents.}
Each entry $\bar z_{ij}$ of $\bar z_i$ represents agent~$i$'s preference for the $j$-th available strategy.
{\color{black} The {\it relative opinion state} $z_{ij} = \bar z_{ij} - \frac{1}{N_s} \sum_{l=1}^{N_s} \bar z_{il}$ defines an agent's preferred strategies, i.e., the inequality $z_{ij} > 0$ can be interpreted as the agent favoring strategy~$j$ relative to other strategies and the magnitude $|z_{ij}|$ denotes the level of its preference.}
 Under  logit choice \eqref{eq:logit_choice}, the higher $\bar z_{ij}$ relative to other entries of $\bar z_i$, the more likely agent~$i$ selects strategy~$j$. 
\eqref{eq:logit_choice} can be interpreted as the best response with respect to the opinion state $\bar z_i$ subject to a random perturbation \cite{HOFBAUER200747}. 

Given  mixed strategy profile $x \in \mathbb X$, we let  $U_i(x) = (U_{i1}(x), \cdots, U_{i N_s}(x)) \in \mathbb R^{N_s}$ be the  \textit{payoff function} for agent~$i$. 
Entry $U_{ij}(x)$ defines agent $i$'s payoff associated with strategy~$j$.
The following are examples of multi-agent games. 
\begin{example} [Prisoner's Dilemma]
  Consider two agents, each with two available strategies: cooperate (strategy~$1$) and defect (strategy~$2$). When both agents cooperate or defect, they receive payoff $p_{CC}$ or $p_{DD}$, respectively. If one defects while the other cooperates, the former receives payoff $p_{DC}$ and the latter receives $p_{CD}$. The payoff function $U_i$ is
  \begin{align} \label{eq:payoff_prisoners_dilemma}
    U_{i} (x) = \begin{pmatrix} U_{i1} (x) \\ U_{i2} (x) \end{pmatrix} = 
    \begin{pmatrix}
      p_{CC} & p_{CD} \\ p_{DC} & p_{DD}
    \end{pmatrix} x_{-i}, ~ i \in \{1, 2\}
  \end{align}
  where, as shorthand notation, we let $x_{-1} = x_2$ and $x_{-2} = x_1$. The parameters $p_{CC}, p_{CD}, p_{DC}, p_{DD}$ satisfy $p_{DC} > p_{CC} > p_{DD} > p_{CD}$, which means that the agents have individual incentives to defect and receive $p_{DD}$, even though they would receive the higher payoff $p_{CC}$ by cooperating.
\end{example}


\begin{example} [Public Goods Game]
  There are $N_a$ agents and $N_s$ strategies. 
  Each agent has a total wealth of ${a (N_s-1)}$ and selects a strategy $j$ in $\{1, \cdots, N_s\}$  that corresponds to contributing $a (N_s-j)$ 
  to a public pool. The total contribution is  multiplied by a factor $\rho$ and  distributed equally among all agents. The payoff function $U_i$ is 
  \begin{multline} \label{eq:payoff_public_goods}
    U_{ij} (x) = a \left( j-1 \right) + \frac{\rho}{N_a} \textstyle\sum_{\substack{k \neq i \\ k=1}}^{N_a} \textstyle\sum_{l=1}^{N_s} a (N_s-l) \, x_{kl} \\ + \frac{\rho}{N_a} a(N_s - j), ~ i \in \{1, \cdots, N_a\}, ~ j \in \{1, \cdots, N_s\},
  \end{multline}
  where $a>0$ and $N_a > \rho > 1$. According to \eqref{eq:payoff_public_goods}, regardless of the others' contributions, each agent  receives the highest payoff when it makes no contribution to the pool. Hence, 
  the rational agent  contributes nothing, i.e., chooses $j=N_s$.
\end{example}


\bigskip 
We define  rate-of-change $\dot{\bar{z}}_i = d\bar{z}_i/dt$ of agent $i$'s opinion state $\bar z_i$
in response to payoffs and social interactions, with 
the continuous-time nonlinear opinion dynamics model 
\cite{Bizyaeva2021}\footnote{\color{black} In $\S$\ref{sec:analysis}, we explain how \eqref{eq:opinion_dynamics_model_zbar} relates to its original form presented in \cite{Bizyaeva2021}. For concise presentation, we omit time dependency of the variables in \eqref{eq:opinion_dynamics_model_zbar}.}: 
\begin{align} \label{eq:opinion_dynamics_model_zbar}
  \dot{\bar z}_{ij} = -d_i \left( \bar z_{ij} - u_i \textstyle\sum_{k=1}^{N_a} 2 R \left( A_{ik}^j  z_{kj} \right) - U_{ij} (x) \right),
\end{align}
with $\bar z_{i}(0) \in \mathbb R^{N_s}$.
\rev{$A_{ik}^j \in \mathbb{R}$ is the weight agent $i$ places in its evaluation of strategy $j$ on its observation of agent $k$'s opinion of strategy $j$.}
The constant \textit{resistance parameter} $d_i>0$ 
reflects the speed with which agent $i$'s opinions change; 
the \textit{attention parameter} $u_i > 0$ 
\rev{reflects the weight placed on incentives derived from social interactions}, where $R: \mathbb R \to [0, 1]$. 
Thus, the state $\bar z_i$ of agent~$i$, and hence its strategy selection, evolves according to the accumulation over time, with the discount factor $d_i$, of the payoffs $U_{ij} (x)$ and 
social incentives $R (A_{ik}^j z_{kj})$. 

We define $R$ as \rev{the saturating function}
\begin{align} \label{eq:social_interaction}
  R (A_{ik}^j  z_{kj}) = \mathbf P \left( A_{ik}^j  z_{kj} \geq \epsilon \right),
\end{align}
where $\epsilon$ is a random variable with a symmetric and unimodal probability density function, e.g., the standard normal distribution. 
To interpret, suppose $A_{ik}^j\geq 0$. 
Then
$A_{ik}^j$ quantifies the influence of noise $\epsilon$ on  inter-agent interactions: the larger $A_{ik}^j$, the smaller the effect of  noise $\epsilon$.\footnote{\color{black} See \S\ref{sec:remarks} for more discussions on the parameter $A^{j}_{ik}$.} Thus, we can interpret \eqref{eq:social_interaction} as a probabilistic model of agent $i$'s perception  of agent~$k$'s preference for strategy~$j$ over other strategies.

\subsection{Emergence of Cooperative Equilibrium} \label{sec:equilibrium_analysis}
In this section, using the prisoner's dilemma as an illustrative example, \rev{we provide intuition} for  how the equilibria of \eqref{eq:opinion_dynamics_model_zbar} depend on system parameters, and  under what parameter regime a cooperative equilibrium emerges.
To simplify the presentation, let $d_i = d$, $u_i = u$,  $A_{ii}^j = \alpha$, and $A_{ik}^j = \gamma$ if $i \neq k$. Let $\bar z^\ast$ be an equilibrium  of \eqref{eq:opinion_dynamics_model_zbar} that satisfies 
\begin{align} \label{eq:equilibrium_condition}
  \bar z_{ij}^\ast &= 2u \left( R (\alpha z_{ij}^\ast) + \textstyle\sum_{\substack{k \neq i\\k=1}}^{N_a} R (\gamma z_{kj}^\ast) \right) + U_{ij} (x^\ast),
\end{align}
where $x_{ij}^\ast = \sigma_j (z_i^\ast)$.

Note that by
\eqref{eq:social_interaction}, 
in a dense subset of the tangent space $T\mathbb X$ of $\mathbb X$, as the influence of the noise in the social interaction becomes arbitrarily small, i.e., $\alpha, \gamma$ are arbitrarily large, $R(\gamma z_{kj})$ converges to a binary ($\{0, 1\}$-valued) function.
If $\alpha, \gamma$ are sufficiently large, we can approximate \eqref{eq:equilibrium_condition} as 
  $\bar z_{ij}^\ast \approx 2u \, n_j^\ast + U_{ij} (x^\ast)$,
where $n_j^\ast$ is the number of agents $k$ having a positive 
opinion $z_{kj}^\ast$ of strategy~$j$ at  equilibrium.
As the attention  $u$ increases, each agent tends to favor the most popular strategy even though selecting other strategies would return higher payoffs.
It follows that the social interaction $R$ incentivizes each agent to reciprocate with other agents in the strategy selection, and the level of reciprocation is determined by the attention parameter $u$ and the number $n_j^\ast$ of agents preferring the same strategy under consideration.

\textit{Example:}
With two reciprocating agents ($N_a=2$, $\alpha = 0, \gamma > 0$) playing the prisoner's dilemma ($N_s = 2$), the equilibrium  $\bar z^\ast$ satisfies $\bar z_{i1}^\ast \approx 2u \, n_1^\ast + U_{i1} (x^\ast)$, where $n_1^\ast \in \{0, 1\}$ indicates whether the opponent 
cooperates ($n_1^\ast=1$) or defects ($n_1^\ast = 0$). If the attention parameter satisfies $2u > \max(p_{DC}-p_{CC}, p_{DD}-p_{CD})$, then for sufficiently large $\gamma$, cooperation becomes an equilibrium of \eqref{eq:opinion_dynamics_model_zbar}. Moreover, given any arbitrarily large $u$, there is a minimum value of $\gamma$ below which cooperation will not be an equilibrium. 

\subsection{Rationality and Reciprocity in the Model
} \label{sec:rationality_reciprocity}
In this section we show how  the model \eqref{eq:opinion_dynamics_model_zbar} captures a range of features observed in human decision-making, including (bounded) rationality \cite{10.2307/2729218} and reciprocity \cite{RePEc:eee:eecrev:v:42:y:1998:i:3-5:p:845-859, 10.2307/2092623}.
We begin by showing that \eqref{eq:opinion_dynamics_model_zbar} generalizes the exponentially discounted reinforcement learning (EXP-D-RL) model studied in \cite{9022871} where every agent makes an individually rational decision by selecting payoff-maximizing strategies. 
To see this, let ${A_{ik}^j = 0}$ for $i,k \in \{1, \cdots, N_a\}$ and $j \in \{1, \cdots, N_s\}$ for which
the social interaction $R(A_{ik}^j z_{kj})$ becomes constant, i.e., $R (A_{ik}^j z_{kj}) = 0.5, ~ \forall z_{kj} \in \mathbb R$. By translating 
$\bar z_{ij}$ by constant $u_i N_a$ and since the logit choice function is invariant with respect to the translation of $\bar z_{ij}$, 
\eqref{eq:opinion_dynamics_model_zbar} specializes to
\begin{align*}
  \dot{\bar z}_{ij} = -d_i \left( \bar z_{ij} - U_{ij} (x) \right), \;\;\; 
  x_{ij} = \frac{\exp\left( \eta^{-1} \bar z_{ij} \right)}{\sum_{l=1}^{N_s} \exp\left( \eta^{-1} \bar z_{il} \right)},
\end{align*}
which is the EXP-D-RL model presented in \cite{9022871}. In this sense, our model \eqref{eq:opinion_dynamics_model_zbar} realizes rationality.

To discuss reciprocity of the opinion dynamics, we consider a two-agent $(N_a=2)$ two-strategy $(N_s=2)$ case.
Suppose that $A_{ik}^j = \eta^{-1}$ if $i \neq k$ and $A_{ik}^j = 0$ otherwise, where $\eta$ is the noise level constant in the logit choice function \eqref{eq:logit_choice}. Then, with 
$R(\cdot) = (\tanh(\cdot) + 1)/2$, we have $R(A_{ik}^j z_{kj}) = x_{kj}$ if $i \neq k$ and $R(A_{ik}^j z_{kj}) = 0.5$ otherwise.

For small $h>0$, assuming that $U_{ij}$ is arbitrarily small, we can approximate the opinion dynamics model \eqref{eq:opinion_dynamics_model_zbar} as
\begin{align*} 
  \bar z_{ij}(t+h) - \bar z_{ij}(t) \approx - h d_i \left( \bar z_{ij}(t) - 2 u_i  x_{-ij}(t) \right).
\end{align*}
For sufficiently large $d_i$, by evaluating the opinion state at time instant $t+h$ with $h = d_i^{-1}$, we observe that
\begin{align} \label{eq:opinion_dynamics_model_approx}
  \bar z_{ij}(t+h) \approx  2 u_i x_{-ij}(t).
\end{align}
Recall that $x_{-ij}$ is the $j$-th entry of the mixed strategy $x_{-i}$ of the opponent of agent~$i$. According to \eqref{eq:opinion_dynamics_model_approx}, with large $u_i$, it holds that $x_{ij} (t+h) = 1$ if and only if $x_{-ij}(t) = 1$. In the prisoner's dilemma, under \eqref{eq:opinion_dynamics_model_approx}, each agent~$i$ decides to cooperate (or defect) if its opponent does so at the previous stage. This behavior resembles TFT, a well-known reciprocity-based strategy in discrete-time iterated games \cite{Axelrod84}. In this sense, our model \eqref{eq:opinion_dynamics_model_zbar} realizes reciprocity. 

\subsection{Further Remarks on the Model \eqref{eq:opinion_dynamics_model_zbar}} \label{sec:remarks}


{\bf Social interaction encourages reciprocity: }
When $A_{ik}^j >0$ for $i \neq k$, the social interaction in \eqref{eq:opinion_dynamics_model_zbar} encourages reciprocity 
by incentivizing each agent 
to select the strategies preferred by other agents. As shown in \S\ref{sec:simulations}, in the prisoner's dilemma and public goods game, such a social interaction mechanism leads to decision-making  representative of human behavior; notably, the agents conditionally cooperate. 
This contrasts with the outcomes of 
rationality-based models 
where agents fail to cooperate (or coordinate).


Our model and analysis can be readily extended to a more general case, as in \cite{Bizyaeva2021}, where the social interaction term in \eqref{eq:opinion_dynamics_model_zbar} is given by
$u_i \sum_{k=1}^{N_a} \sum_{l=1}^{N_s} 2R \left( A_{ik}^{jl} z_{kl} \right)$. 
In this generalization, agent  $i$'s opinion of strategy $j$ may also depend on other agent opinions  of strategies $l \neq j$.

{\bf Network structure: }
The  $A_{ik}^j$ in 
\eqref{eq:opinion_dynamics_model_zbar} define a network structure among agents for  strategy $j$. 
One can specify the presence ($A_{ik}^j > 0$ for reciprocal,  $A_{ik}^j<0$ for antagonistic) or lack ($A_{ik}^j = 0$) of interaction between agents~$i$ and $k$ in their selecting strategy~$j$. 
We prove results on the role of network structure in our model in \S\ref{sec:analysis}. See \cite{Bizyaeva2021,Franci2021} for more on network structure and the nonlinear opinion dynamics.

{\color{black}
\section{Bistability Analysis of 2-Strategy Games} \label{sec:analysis}
We present bistability analysis for \eqref{eq:opinion_dynamics_model_zbar} in two-strategy games with homogeneous parameters.\footnote{The proofs of all the theorems are provided in the Appendix.} We assume 
$G = (\mathbb V, \mathbb E)$ and $\hat{G} = (\mathbb V, \hat{\mathbb E})$, with $\mathbb V = \{1, \cdots, N_a\}$, are simple graphs governing the social interaction and game interaction, respectively, and $A = (a_{ik})_{i,k \in \mathbb V}$ and $\hat A = (\hat a_{ik})_{i,k \in \mathbb V}$ are the corresponding adjacency matrices. We assume  the payoff function has the  form:
\begin{align} \label{eq:homogeneous_payoff_function}
\begin{pmatrix}
  U_{i1}(x) \\
  U_{i2}(x)
\end{pmatrix}
= \textstyle\sum_{k \in \hat{\mathbb E}} \begin{pmatrix} p_{11} & p_{12} \\ p_{21} & p_{22} \end{pmatrix} x_k + \begin{pmatrix} b_1 \\ b_2 \end{pmatrix},
\end{align}
and the parameters of \eqref{eq:opinion_dynamics_model_zbar} are given by
$d_i = d$, $u_i=u$, and $A_{ii}^j = \alpha > 0$, and $A_{ik}^j = \gamma a_{ik} \geq 0$  if $i \neq k$. 

For analysis, we adopt the original form of  \eqref{eq:opinion_dynamics_model_zbar} from \cite{Bizyaeva2021}:
\begin{align} \label{eq:opinion_dynamics_model}
  &\dot z_{ij} = F_{ij} \left( z \right) - \frac{1}{N_s} \textstyle\sum_{l=1}^{N_s} F_{il} \left(  z \right), ~ \textstyle\sum_{j=1}^{N_s} z_{ij}(0) = 0, \\
  &F_{ij}(z) = -d \left( z_{ij} \! - \! u \left( S \left( \alpha z_{ij} \right) \! + \! \textstyle\sum_{k \in \mathbb E} S \left( \gamma z_{kj} \right) \right) - U_{ij} (x) \right) \nonumber
\end{align}
where $z_{ij}(0) = \bar z_{ij}(0) - \frac{1}{N_s} \sum_{l=1}^{N_s} \bar z_{il}(0)$ and the saturation function $S$ is given by $S = 2R-1$.
The variable $z = (z_1, \cdots, z_{N_a}) \in T\mathbb X$ denotes the relative opinion state. 
In Theorem~\ref{thm:model_equivalence}, we show that models \eqref{eq:opinion_dynamics_model_zbar} and \eqref{eq:opinion_dynamics_model} are related by the projection $z_i = P_0 \bar z_i$, where $P_0 = I - \frac{1}{N_s} \mathbf 1 \mathbf 1^T$, and yield the same transient and steady-state mixed-strategy behavior.

 \begin{theorem} \label{thm:model_equivalence}
The following two statements are true.

\textbf{i)} If $\bar z(t)$ is a solution of \eqref{eq:opinion_dynamics_model_zbar}, then $z(t)$, satisfying $z_i(t)=P_0 \bar z_i(t)$, is a solution of \eqref{eq:opinion_dynamics_model}. Conversely, if $z(t)$ is a solution of \eqref{eq:opinion_dynamics_model}, then $\bar z(t)$ defined as
$\bar z_{ij}(t) = e^{-d t} \bar z_{ij}(0)+ d \int_0^t e^{-d (t-\tau)} ( 2u ( R ( \alpha z_{ij} (\tau)) +  \textstyle\sum_{k \in \mathbb E}
  R ( \gamma z_{kj} (\tau)) ) + U_{ij} (x(\tau)) ) \, \mathrm d \tau$ with $x_{ij} = \sigma_j (z_i)$ satisfies $z_i(t)=P_0 \bar z_i(t)$ and is a solution of \eqref{eq:opinion_dynamics_model_zbar}.

\textbf{ii)} If $\bar z^\ast$ is a stable (unstable) equilibrium of \eqref{eq:opinion_dynamics_model_zbar}, then $z^\ast$, satisfying $z_i^\ast=P_0\bar z_i^\ast$, is a stable (unstable) equilibrium of \eqref{eq:opinion_dynamics_model}. Conversely, if $z^*$ is a stable (unstable) equilibrium of \eqref{eq:opinion_dynamics_model} then $\bar z^*$, defined as $\bar z^*_{ij}=2u ( R ( \alpha z_{ij}^\ast ) +  \textstyle\sum_{k \in \mathbb E }
  R ( \gamma z_{kj}^\ast ) )+U_{ij}(x^\ast)$ with $x_{ij}^\ast = \sigma_j(z_i^\ast)$
satisfies $z_i^\ast=P_0 \bar z_i^\ast$ and is a stable (unstable) equilibrium of \eqref{eq:opinion_dynamics_model_zbar}.
\end{theorem}



We further assume that $S$ satisfies the following conditions: $S$ is odd sigmoidal and it holds that $S(0)=0$, $S' (0)>0 $, $\mathrm{sign} \, S''(a) = - \mathrm{sign}(a), ~ \forall a \in \mathbb R$, and $S'''(0) =-2$.\footnote{To simplify the notation, without loss of generality, we make the assumption that $S'''(0) =-2$, for instance, by scaling $S$.} 
Since  $z_{i1} = - z_{i2}$, we can simplify the expression \eqref{eq:opinion_dynamics_model} as
\begin{multline} \label{eq:reduced_vector_field}
  \dot{\mathbf{z}}
  = -d  \Big(\mathbf{z} - u \left( S(\alpha \mathbf{z}) + A S(\gamma \mathbf{z}) \right) \\
   - \frac{1}{4} p \hat{A} \tanh(\eta^{-1} \mathbf{z}) - \frac{1}{4} p^{\perp} \hat{A} \mathbf{1}   - (b_1 - b_2) \mathbf{1} \Big) 
\end{multline}
with $\mathbf{z} = (z_{11}, \cdots, z_{N_a 1})$, 
$p = p_{11} - p_{12} - p_{21} + p_{22}$, $p^{\perp} = p_{11} + p_{12} - p_{21} - p_{22}$, and $S(\gamma \mathbf z) = (S(\gamma z_{11}), \cdots, S(\gamma z_{N_a 1}))$.

\begin{figure}
  \center
  \subfigure[$p_{CC}=15$]{
    \includegraphics[trim={.0in .1in 0in .1in}, clip, height=.72in, page=1]{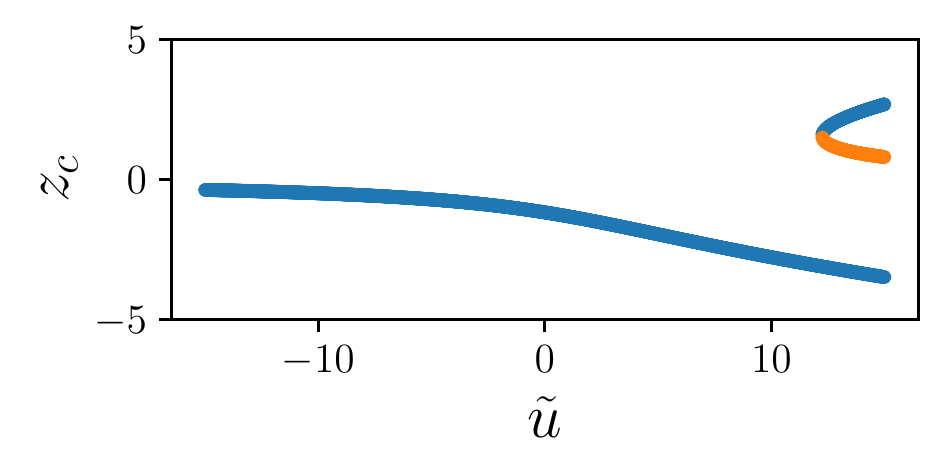}
    \label{fig:bifurcation_diagram_pd_a}
  }
  \subfigure[$p_{CC}=30$]{
    \includegraphics[trim={.32in .1in 0in .1in}, clip, height=.72in, page=2]{./figures/bifurcation_diagram_pd.pdf}
    \label{fig:bifurcation_diagram_pd_b}
  }
  
  \subfigure[$a=40$]{
    \includegraphics[trim={.0in .1in 0in .1in}, clip, height=.66in, page=3]{./figures/bifurcation_diagram_pg.pdf}
    \label{fig:bifurcation_diagram_pg_a}
  }
  \subfigure[$a=10$]{
    \includegraphics[trim={.32in .1in 0in .1in}, clip, height=.66in, page=1]{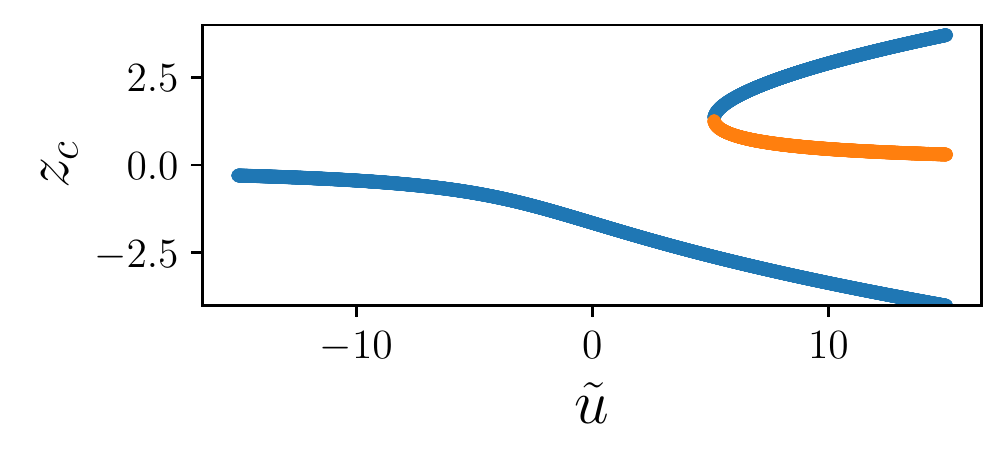}
    \label{fig:bifurcation_diagram_pg_b}
  }
  
  \caption{Bifurcation diagram of 
  \eqref{eq:lyapunov_schmidt_reduction}
  in the prisoner's dilemma with (a) $p_{CC}=15$ and (b) $p_{CC}=30$, where $\alpha = 0$, $d=\gamma=\eta=1$, $p_{CD}=0$, $p_{DC}=40$,  $p_{DD}=5$, and in the public goods game with (c) $a=40$ and (d) $a=10$, where $\alpha = 0$, $d=\gamma=\eta=1$, $\rho=2$, and $N_a=20$. Blue (orange) curves are  stable (unstable) equilibrium states. Solutions with $z_c<0$ ($z_c>0$) correspond to mutual defection (cooperation). Mutual defection is always stable; 
  for large enough $\tilde u$, mutual cooperation is stable. 
  }
  \label{fig:bifurcation_diagram}
  \vspace{-3ex}
\end{figure}

\begin{theorem} [Bistability in games] \label{theorem:bistability}
 Consider \eqref{eq:reduced_vector_field}. Let $\zeta_{max}(u,\gamma,p)$ be the largest-real-part  eigenvalue  of $u \gamma S'(0)A + \frac{1}{4} \eta^{-1} p \hat{A}$ and $v_{max}$ $(w_{max})$ be its corresponding right (left) eigenvector.
 
 \textbf{i)}  Suppose $\zeta_{max}$ is real and simple, and $w_{max}^T \gamma S'(0) A v_{max} > 0$ holds.
 When $p^\perp = b_1 = b_2 = 0$, there exists a critical value $u^\ast$ for which if $u < u^\ast$, the origin $\mathbf{z} = 0$ is locally exponentially stable, and if $u > u^\ast$, the origin is unstable and two bistable equilibrium solution branches emerge in  a symmetric pitchfork bifurcation along a manifold tangent to the span of $v_{max}$. When $p^\perp$, $b_{1}$, and/or $b_{2}$ are nonzero, the system is an unfolding of the symmetric pitchfork bifurcation, and the parameter
 \begin{align} \label{eq:unfolding_parameter}
     b = \langle w_{max}, \frac{1}{4} d p^{\perp} \hat{A} \mathbf{1} + d (b_1 - b_2) \mathbf{1} \rangle
 \end{align}
 determines the direction of the unfolding. Furthermore, $u^\ast$ depends on $p$ according to $\frac{\partial u^*}{ \partial p } = - \frac{1}{4 \alpha S'(0)} w_{max}^T \hat{A} v_{max}$. 
 
 \textbf{ii)} Suppose $ u \gamma S'(0) A + \frac{1}{4} \eta^{-1} p \hat{A}$ is an irreducible nonnegative matrix.\footnote{This 
 holds, e.g., when $\gamma > 0$, $p \geq 0$, and at least one of $A$, $\hat{A}$ corresponds to a connected graph.} Near $u^*$, for the bistable equilibria, $\operatorname{sign}(z_{i1}) = \operatorname{sign}(z_{k1})$, $\forall i,k$, i.e., all agents favor the same strategy. 
 
 \textbf{iii)} 
 Suppose $v_{max}$ ($w_{max}$) is also left (right) eigenvector of both $A$ and $\hat{A}$. 
 Denote by $\lambda$, $\hat{\lambda}$ the eigenvalues of $A$, $\hat{A}$, respectively,  corresponding to $v_{max} (w_{max})$. 
 Then $u^* = \frac{1 - \frac{1}{4} \eta^{-1} p \hat{\lambda}}{S'(0) (\alpha + \gamma \lambda )}$, and the  unfolding parameter \eqref{eq:unfolding_parameter} simplifies to 
 \begin{equation} \label{eq:unfolding_parameter_simplified}
    b = d \left(\frac{1}{4} p^{\perp} \hat{\lambda} + b_1 - b_2\right) \langle w_{max}, \mathbf{1} \rangle .
 \end{equation}
 
\end{theorem}

The following theorem shows how the bifurcation depends on degree (number of neighbors) for  $G, \hat G$  regular graphs.
\begin{theorem} \label{thm:Kreg_graphs}
 Suppose  $\gamma > 0$, $p \geq 0$, and 
 $G$, $\hat G$ are undirected, connected, and regular with degrees $K$, $\hat K$, respectively.
 The bifurcation point $u^\ast$ and unfolding parameter $b$ satisfy
 $\operatorname{sign} \left( \frac{\partial u^*}{\partial K} \right) = \operatorname{sign}\left(\frac{1}{4} \eta^{-1} p \hat{K}-1\right) $, $\operatorname{sign} \left( \frac{\partial u^*}{\partial \hat{K}} \right) = \operatorname{sign}(-p)$, and $\operatorname{sign} \left(\frac{\partial b}{\partial \hat{K}} \right) = \operatorname{sign}(p^\perp)$. 
\end{theorem}

\begin{remark}
For games with more than $2$ strategies and heterogeneous payoff functions, the analysis can be generalized using 
$
U_{ij}(x) = \textstyle\sum_{k \in \hat{\mathbb E}} \begin{pmatrix} p_{j1}^{ik} & \hdots & p_{j N_s}^{ik} \end{pmatrix} x_k  + b_j^i.
$
\end{remark}



In what follows, we discuss implications of Theorems~\ref{theorem:bistability},\ref{thm:Kreg_graphs} in social dilemmas using the prisoner's dilemma and public goods game. From now on, we take $S(\cdot) = \tanh(\cdot)$.


\textbf{Prisoner's dilemma: }
Let $p_{11}=p_{CC}$, $p_{12}=p_{CD}$, $p_{21}=p_{DC}$, $p_{22}=p_{DD}$ and $b_1=b_2=0$ so \eqref{eq:homogeneous_payoff_function} specializes to \eqref{eq:payoff_prisoners_dilemma}.
\begin{corollary} \label{cor:prisoners_dilemma}
For $\gamma > 0$, the following hold: $v_{max} = w_{max} = (1,1)$, $\lambda = \hat{\lambda} = 1$, and $p^{\perp} < 0$. Hence, we have $u^{*} = \frac{1 - \frac{1}{4}\eta^{-1}p}{\alpha + \gamma}$ and $b = \frac{1}{2} d p^{\perp} < 0$.
\end{corollary}

Figs.~\ref{fig:bifurcation_diagram_pd_a},\ref{fig:bifurcation_diagram_pd_b} show the bifurcation diagram (plot of equilibria as a function of bifurcation parameter $\tilde u = u - u^*$) of the Lyapunov-Schmidt reduction \eqref{eq:lyapunov_schmidt_reduction} of \eqref{eq:opinion_dynamics_model},  for two  values of $p_{CC}$.
$p$ and $p^\perp$ have a two-fold effect: \textit{i)} $p$ changes the location of the pitchfork bifurcation point in the $\tilde u$-axis. 
\textit{ii)} Since $p^{\perp} < 0$, the pitchfork bifurcation unfolds favoring the branch of solutions corresponding to mutual defection.
For sufficiently large $\tilde u$ (equivalently, $u$), a branch of solutions corresponding to mutual cooperation emerges, 
and the larger the $u$ the larger its region of attraction. A small  $u$ is required for larger $p_{CC}$, since it decreases incentive to defect.

If $p_{11} >p_{21}=40$,  
the game is the Stag Hunt where the strategy to hunt a stag replaces cooperation and the strategy to hunt a hare replaces defection. 
Coordinated stag hunting and coordinated hare hunting are both Nash equilibria, the former  payoff-dominating and the latter risk-dominating. The model predicts the larger the $u$, the larger the region of attraction to coordinated stag hunting. 

\textbf{Public goods game: }
Let $N_s = 2$, i.e., each agent decides to cooperate and contributes its entire wealth $(j=1)$, or defect and contributes nothing $(j=2)$. 
Note that \eqref{eq:homogeneous_payoff_function} specializes to \eqref{eq:payoff_public_goods} by selecting $p_{11} = p_{21} = b_1 = a \rho / N_a$, $p_{12}=p_{22}=0$, and $b_2=a$ with all-to-all graph $\hat G$. 

\begin{corollary} \label{cor:public_goods}
With $N_s=2$,
it holds that $p = p^{\perp} = 0$. For $\gamma > 0$ and connected graph $G$, the following hold: \textit{i)} The eigenvectors $v_{max}, w_{max}$ have all nonzero same-sign entries, and 
$u^* = \frac{1}{\alpha + \gamma \lambda}$ and
$b = - d a(1 - \rho/N_a) \langle w_{max},\mathbf{1} \rangle < 0$. \textit{ii)} When $A$ is regular with degree $K$, it holds that $\frac{\partial u^*}{\partial K} < 0$, i.e., with larger $K$, bistability requires less attention $u$.
\end{corollary}

Figs.~\ref{fig:bifurcation_diagram_pg_a},\ref{fig:bifurcation_diagram_pg_b} show the bifurcation diagram
for two  values of  $a$. 
Since $p=0$, it has no effect. 
However, $b_1 - b_2 = -a(1- \rho/N_a)<0$; hence, for reciprocating agents ($\gamma > 0$), the pitchfork bifurcation unfolds towards the branch of solutions corresponding to no agent contributing to the public pool. Since the strength of the unfolding is proportional to $a$, emergence of the mutually cooperative solution, when all agents contribute, requires a smaller $u$ for smaller $a$ and for a fixed $u$ its region of attraction grows as $a$ decreases.} 




\section{Numerical Studies} \label{sec:simulations}
\begin{figure} [t]
  \center
  \subfigure[$r = 20$]{
    \includegraphics[trim={.1in .1in .45in .1in}, clip, height=.9in, page=1]{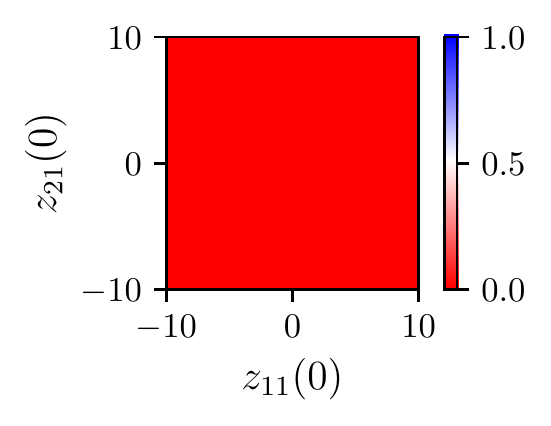}
    \label{fig:pd_simulation_01_a}
  }~
  \subfigure[$r = 10$]{
    \includegraphics[trim={.35in .1in .45in .1in}, clip, height=.9in, page=1]{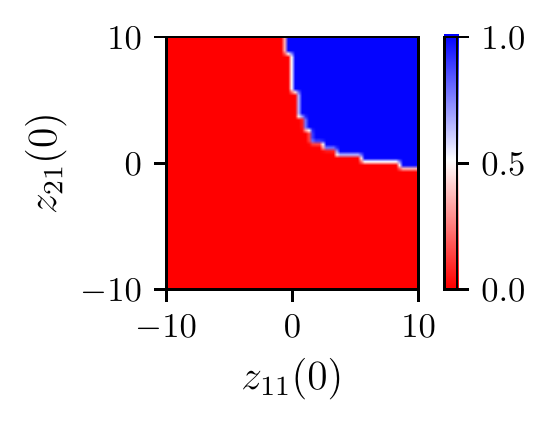}
    \label{fig:pd_simulation_01_b}
  }~
  \subfigure[$r = 0$]{
    \includegraphics[trim={.35in .1in .1in .1in}, clip, height=.9in, page=1]{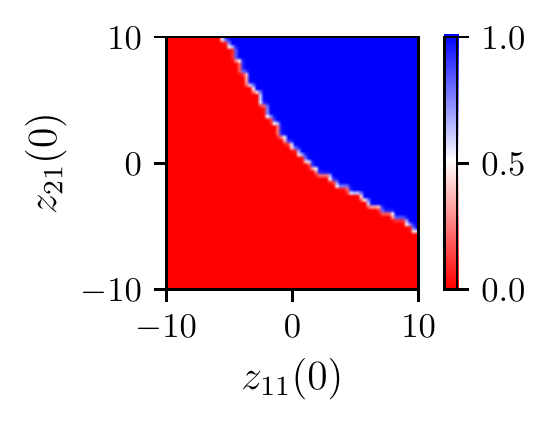}
    \label{fig:pd_simulation_01_c}
  }
  
  \caption{Heatmaps depict the probability of mutual cooperation in the prisoner's dilemma for three different values of $r$ in the payoff matrix \eqref{eq:payoff_matrix_simulation}.}
  \label{fig:pd_simulation_01}
 \vspace{-2.0ex}
\end{figure}

\begin{figure} [t]
  \center
  \subfigure[$a = 40$]{
    \includegraphics[trim={.1in .1in .1in .1in}, clip, height=.88in, page=1]{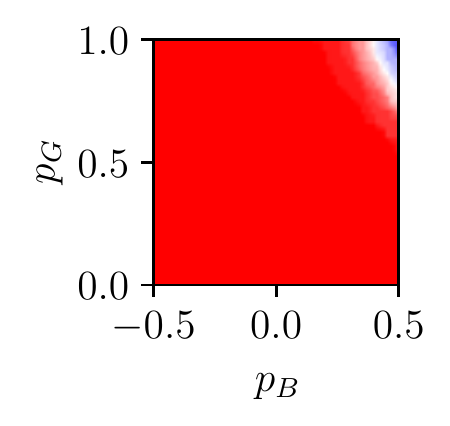}
    \label{fig:pg_simulation_01_a}
  }~
  \subfigure[$a = 20$]{
    \includegraphics[trim={.3in .1in .1in .1in}, clip, height=.88in, page=1]{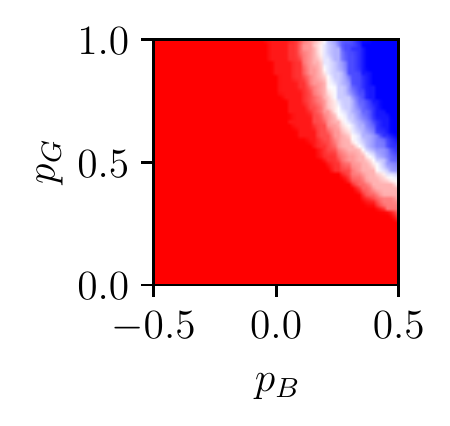}
    \label{fig:pg_simulation_01_b}
  }~
  \subfigure[$a = 10$]{
    \includegraphics[trim={.3in .1in .1in .1in}, clip, height=.88in, page=1]{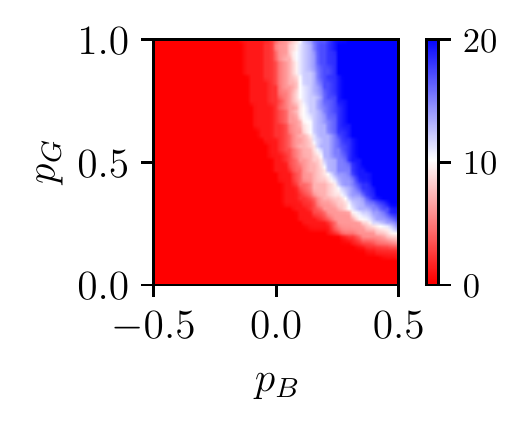}
    \label{fig:pg_simulation_01_c}
  }
  
  \caption{Heatmaps depict the average number of agents cooperating in the public goods game for three different values of $a$ in the payoff matrix \eqref{eq:payoff_public_goods_simulation}.}
  \label{fig:pg_simulation_01}
 \vspace{-4ex}
\end{figure}

\subsection{Prisoner's Dilemma} \label{sec:simulations_pd}
We set $d_i = \eta = 1$, $u_i = 10$,  $A_{ii}^j = 0$, 
$A_{ik}^j = 1$ for $i \neq k$, and $S(\cdot) = \tanh (\cdot)$.
Consider the payoff matrix \eqref{eq:payoff_prisoners_dilemma} given by
\begin{align} \label{eq:payoff_matrix_simulation}
  \begin{pmatrix}
    p_{CC} & p_{CD} \\ p_{DC} & p_{DD}
  \end{pmatrix} = \begin{pmatrix}
    35 & 0-r \\ 40+r & 5
  \end{pmatrix}
\end{align}
and $r>0$ is an extra reward (penalty) an agent receives if it defects (cooperates) while its opponent cooperates (defects). 

Using simulations, we illustrate limit points of the opinion state trajectories, predicted by the theory.
In Fig.~\ref{fig:pd_simulation_01}
each heatmap illustrates the probability of both agents cooperating and the two axes represent the initial opinion states $z_{11}(0), z_{21}(0)$ of the agents associated with the cooperation strategy. Since the two agents are
reciprocating, for all cases, we observe that the heatmaps for both agents are identical, and hence we only present that of agent 1.
 
In Figs.~\ref{fig:pd_simulation_01_b} and \ref{fig:pd_simulation_01_c}, we can observe that when both agents are \textit{nice}, i.e., the agents' initial opinion states $z_{11}(0)>0, z_{21}(0)>0$ for the cooperation are large enough, they can maintain mutual cooperation. Also, a sufficiently nice agent ($z_{i1}(0)>0$) \textit{forgives}  the exploiting behavior (defection) of its opponent that initially is not nice ($z_{-i1}(0)<0$). However, when its opponent has a strong intention to defect ($z_{-i1}(0)$ substantially large), the agent also defects to avoid being exploited and is \textit{provocable}.

An increase in $r$ motivates the agents to defect (Fig.~\ref{fig:pd_simulation_01}). When $r=20$, since $p_{DC}-p_{CC} =  p_{DD}-p_{CD} = 25 > 2u = 20$, the cooperation strategy is dominated by the defection strategy, and  both agents eventually defect (Fig.~\ref{fig:pd_simulation_01_a}). 
Thus, as predicted by the theory and illustrated in Figs.~\ref{fig:bifurcation_diagram_pd_a},\ref{fig:bifurcation_diagram_pd_b}, when there is a strong enough incentive to defect, the level of attention $u$ to opinion exchanges, which translates into the level of reciprocity, may be insufficient to prevent the agents from pursuing individually rational decision-making. 

{\color{black}
\subsection{Public Goods Game}
For the $2$-strategy public goods game, we adopt the same parameters of \eqref{eq:opinion_dynamics_model} as in \S\ref{sec:simulations_pd} except that $u_i=5$, the inter-agent interactions are governed by the Erd{\"{o}}s-R{\'e}nyi graph with parameter $p_G$ (for $i \neq k$, $A_{ik}^j = 1$ with probability $p_G$ and $A_{ik}^j = 0$ with probability $1-p_G$), and the initial opinion state of each agent is uniformly randomly selected as $z_{i1}(0) \sim \mathrm{Uniform} (-0.5+p_B, 0.5+p_B)$, where $p_B$ is a bias in favor of  cooperation. Let $\rho = 2$, $N_a = 20$ so 
\eqref{eq:payoff_public_goods} is
\begin{align} \label{eq:payoff_public_goods_simulation}
U_{ij} (x) 
= \begin{cases} 
\frac{a}{10} + \frac{a}{10} \textstyle\sum_{\substack{k \neq i \\ k=1}}^{20} x_{k1} & \text{if $j=1$} \\
a + \frac{a}{10}\textstyle\sum_{\substack{k \neq i \\ k=1}}^{20} x_{k1}  & \text{if $j=2$.}
\end{cases}
\end{align}
We evaluate opinion state trajectories over a range  of values of $p_G$, $p_B$, and $a$ to explore how the network structure of the social interaction, initial opinion states, and total wealth tune the emergence of cooperation as predicted by the theory.

Each heatmap in Fig.~\ref{fig:pg_simulation_01} depicts, for a given $a$, the average number of agents that cooperate at steady-state 
for a range of $p_G, p_B$.
Both network structure, 
determined by $p_G$, and  agents' initial preference to contribute to the public pool, determined by $p_B$, play important roles: The cooperation among the $20$ agents is more likely to be sustained if each agent has a greater chance to interact with others ($p_G$ large) and favors  cooperation at the beginning of the game ($p_B$ large). Interestingly, even if they prefer to cooperate at the beginning ($p_B$ large), when the agents are interacting less and cannot perceive the opinion state of others ($p_G$ small), they decide to defect over time. The advantage of large $p_G$ is as for large $K$ for regular graphs, as predicted by Corollary~\ref{cor:public_goods}.

The payoff difference $U_{i2}(x) - U_{i1}(x) = 0.9 a$ between the two strategies depends on the total wealth $a$ and quantifies the incentive for the agents to defect. Consequently, the more wealth agents have, the higher incentive they receive  to not contribute.
This is illustrated in Fig.~\ref{fig:pg_simulation_01}, where mutual defection (cooperation) is more (less) likely as $a$ increases.
}

\section{Final Remarks} \label{sec:conclusion}

We have shown that the nonlinear opinion dynamics model of \cite{Bizyaeva2021,Franci2021} provides an analytically tractable framework for  studying cooperative behavior in repeated multi-agent games, where agents rely on rationality {\em and} reciprocity, both of which are 
central to human decision-making.
\rev{The opinion update depends on a saturated function of inter-agent opinion exchanges, which allows  
mutual cooperation (or coordination) to emerge as one of two bistable equilibria in two-strategy games.} 
For the prisoner's dilemma \rev{and multi-agent public goods game}, mutual cooperation emerges  when the attention $u$ to social interaction, and thus reciprocity, is sufficiently strong.
\rev{The bistability provides a possible mathematical account for how reciprocity enables stable cooperative behavior, as observed in experimental studies, and a principled approach for tuning cooperative behavior.}

Building on coupled opinion-attention dynamic analysis of \cite{Bizyaeva2021,Franci2021}, we will design feedback dynamics for $u$ 
to reflect, for instance, agents' growing appreciation of 
social interactions. This will allow opportunities to influence behavior, e.g., to elicit cooperation or coordination among agents.  
We will also leverage the versatility of the model to investigate games with more than two strategies and heterogeneity.  

\appendix
{\color{black}
\noindent {\em Proof of Theorem~\ref{thm:model_equivalence}:}
\textbf{i)}
The first statement is verified by comparing \eqref{eq:opinion_dynamics_model_zbar} and \eqref{eq:opinion_dynamics_model}. For the second statement, 
by the definition of  $S$ and $z_{ij}(0) = \bar z_{ij}(0) - \frac{1}{N_s} \sum_{l=1}^{N_s} \bar z_{il}(0)$ we get 
$
    \frac{\mathrm d}{\mathrm dt} \left( \bar z_{ij}(t) - \frac{1}{N_s} \sum_{l=1}^{N_s} \bar z_{il}(t) \right) = F_{ij}(z) - \frac{1}{N_s}\sum_{l=1}^{N_s} F_{il}(z).
$
Therefore, $\bar z_{ij}(t) - \frac{1}{N_s} \sum_{l=1}^{N_s} \bar z_{il}(t)$ is a solution of \eqref{eq:opinion_dynamics_model} and hence $z_{ij}(t) = \bar z_{ij}(t) - \frac{1}{N_s} \sum_{l=1}^{N_s} \bar z_{il}(t)$. Thus,  $\sigma_j (z_i (t)) = \sigma_j (\bar z_i (t))$ for all $t \geq 0$ and $\bar z(t)$ is a solution to \eqref{eq:opinion_dynamics_model_zbar}. 

\textbf{ii)} If $\bar z^\ast$ is an equilibrium of \eqref{eq:opinion_dynamics_model_zbar}
then  $z_i^\ast = P_0 \bar z_i^\ast$ satisfies $P_oF_i(z^\ast) = 0$ and hence is an equilibrium of \eqref{eq:opinion_dynamics_model}. 
To prove the second statement, suppose $z^\ast$ is an equilibrium of \eqref{eq:opinion_dynamics_model}. As in the proof for \textbf{i)}, we can establish that $z_{ij}^\ast = \bar z_{ij}^\ast - \frac{1}{N_s} \sum_{l=1}^{N_s} \bar z_{il}^\ast$ for $\bar z_{ij}^\ast$ defined as in the statement. Thus,  $\sigma_j (z_i^\ast) = \sigma_j (\bar z_i^\ast)$ and $\bar z^\ast$ is an equilibrium of \eqref{eq:opinion_dynamics_model_zbar}. The stability of the equilibria follows from \textbf{i)}. \hfill\QED
\vspace{1.5ex}

\noindent {\em Proof of Theorem~\ref{theorem:bistability}:}
\textbf{i)} When $p^{\perp} = b_1 = b_2 = 0$, the neutral state $\mathbf{z} = 0$ is always an equilibrium of \eqref{eq:reduced_vector_field}.  The Jacobian of the linearization of \eqref{eq:reduced_vector_field} at $\mathbf{z} = 0$ is
\begin{equation} \label{eq:2op_Jacobian}
    J(0) = -d\left( (1 \! - \! u S'(0)\alpha) I \!-\! u \gamma S'(0) A \! - \! \frac{1}{4} \eta^{-1} p \hat{A}  \right)
\end{equation}
and its eigenvalues take the form $\mu_{i} = -d (1 - u S'(0) \alpha - \zeta_i(u,\gamma,p))$
where $\zeta_i$ is an eigenvalue of the matrix $u \gamma S'(0) A + \frac{1}{4} \eta^{-1} p \hat{A}$. By \cite{plaut1973derivatives}, we can derive that $\frac{\partial \zeta_{max}}{\partial u} = w_{max}^T \gamma S'(0) A v_{max}$, and
$\frac{\partial \mu_{max}}{\partial u} = d S'(0) \alpha + w_{max}^T \gamma S'(0) A v_{max} > 0$ for any $u, p, \gamma$. Hence, there exists a critical value $u^*$ for which if $u < u^*$, all eigenvalues of \eqref{eq:2op_Jacobian} have negative real part, and if $u>u^*$, $\mu_{max}$ is positive, real, and simple. 
By Lyapunov-Schmidt reduction 
\cite{golubitsky1985}, the one-dimensional dynamics projected onto  span of $v_{max}$ are 
\begin{multline} \label{eq:lyapunov_schmidt_reduction}
    \dot{z}_c = - 2 d \langle w_{max}, 
    \Tilde{v}\rangle z_c^3 + d S'(0) \langle w_{max}, (\alpha I + \gamma A) v_{max} \rangle \\ \times \tilde u z_c  +  \langle w_{max}, \frac{1}{4} d p^{\perp} \hat{A} \mathbf{1} + d (b_1 - b_2) \mathbf{1} \rangle + h.o.t.
\end{multline}
where $\Tilde{v} = v_{max} \odot (\alpha I + \gamma A + \frac{1}{4}p \hat{A}) v_{max} \odot (\alpha I + \gamma A + \frac{1}{4}p \hat{A}) v_{max} $ and
$\tilde u = u - u^*(\alpha,\gamma, p, \eta)$. By the recognition problem \cite[Chapter II, Proposition 9.2]{golubitsky1985}, 
\eqref{eq:lyapunov_schmidt_reduction} describes an unfolding of the pitchfork bifurcation. The last statement follows by implicit differentiation of $-1 + \alpha S'(0) u^* + \zeta_{max} = 0$.

\textbf{ii)} By the Perron-Frobenius theorem, $v_{max}$ and $w_{max}$ have all same-sign entries. The rest follows from part \textbf{i)} and the center manifold theorem.

\textbf{iii)}  By the assumptions on $v_{max}$ ($w_{max}$),  $\zeta_{max} = u \gamma S'(0) \lambda + \frac{1}{4} \eta^{-1} p \hat{\lambda}$, $\mu_{max} = -d (1 - u S'(0) \alpha - u \gamma S'(0) \lambda - \frac{1}{4} \eta^{-1} p \hat{\lambda})$. Thus, $u^* = \frac{1 - \frac{1}{4} \eta^{-1} p \hat{\lambda}}{S'(0) (\alpha + \gamma \lambda )}$. The rest follows from \eqref{eq:unfolding_parameter} since $w_{max}$ is the left eigenvector of $\hat A$. \hfill\QED 
\vspace{1.5ex}

\noindent {\em Proof of Theorem~\ref{thm:Kreg_graphs}:}
By the assumptions on $p$, $G$, and $\hat G$, 
we can verify that $v_{max}$, $w_{max}$, $\lambda$, and $\hat \lambda$, given in Theorem~\ref{theorem:bistability} \textbf{iii)}, satisfy
$v_{max} = w_{max} = \mathbf{1}$, $\lambda = K$, and $\hat{\lambda} = \hat{K}$; then 
$\frac{\partial u^*}{\partial K} = 
    \frac{- \gamma ( 1 - \frac{1}{4} \eta^{-1} p \hat{K})}{S'(0)(\alpha + \gamma K)^2}$ and  
    $\frac{\partial u^*}{\partial \hat{K}} = 
    \frac{- \frac{1}{4} \eta^{-1} p}{S'(0) (\alpha + \gamma K)}$.
From \eqref{eq:unfolding_parameter_simplified}, 
 $\frac{\partial b}{ \partial K} = \frac{1}{4} N_a d p^\perp$, and the theorem follows. \hfill\QED
}

\balance
\bibliographystyle{IEEEtran}
\bibliography{IEEEabrv,reference}

\begin{thebibliography}{10}
\providecommand{\url}[1]{#1}
\csname url@rmstyle\endcsname
\providecommand{\newblock}{\relax}
\providecommand{\bibinfo}[2]{#2}
\providecommand\BIBentrySTDinterwordspacing{\spaceskip=0pt\relax}
\providecommand\BIBentryALTinterwordstretchfactor{4}
\providecommand\BIBentryALTinterwordspacing{\spaceskip=\fontdimen2\font plus
\BIBentryALTinterwordstretchfactor\fontdimen3\font minus
  \fontdimen4\font\relax}
\providecommand\BIBforeignlanguage[2]{{%
\expandafter\ifx\csname l@#1\endcsname\relax
\typeout{** WARNING: IEEEtran.bst: No hyphenation pattern has been}%
\typeout{** loaded for the language `#1'. Using the pattern for}%
\typeout{** the default language instead.}%
\else
\language=\csname l@#1\endcsname
\fi
#2}}

\bibitem{10.2307/2092623}
A.~W. Gouldner, ``The norm of reciprocity: A preliminary statement,''
  \emph{American Sociological Review}, vol.~25, no.~2, pp. 161--178, 1960.

\bibitem{Axelrod84}
R.~Axelrod, \emph{The Evolution of Cooperation}.\hskip 1em plus 0.5em minus
  0.4em\relax Basic Books, 1984.

\bibitem{RePEc:eee:eecrev:v:42:y:1998:i:3-5:p:845-859}
E.~Fehr and S.~Gächter, ``Reciprocity and economics: The economic implications
  of homo reciprocans,'' \emph{European Economic Review}, vol.~42, no. 3-5, pp.
  845--859, 1998.

\bibitem{doi:10.1146/annurev.soc.24.1.183}
P.~Kollock, ``Social dilemmas: The anatomy of cooperation,'' \emph{Annual
  Review of Sociology}, vol.~24, no.~1, pp. 183--214, 1998.

\bibitem{doi:10.1098/rsos.182142}
L.~Heuer and A.~Orland, ``Cooperation in the prisoner's dilemma: An
  experimental comparison between pure and mixed strategies,'' \emph{Royal
  Society Open Science}, vol.~6, no.~7, p. 182142, 2019.

\bibitem{Mao2017}
A.~Mao, L.~Dworkin, S.~Suri, and D.~J. Watts, ``Resilient cooperators stabilize
  long-run cooperation in the finitely repeated prisoner's dilemma,''
  \emph{Nature Communications}, vol.~8, no.~1, p. 13800, 2017.

\bibitem{Bizyaeva2021}
A.~Bizyaeva, A.~Franci, and N.~Leonard, ``Nonlinear opinion dynamics with
  tunable sensitivity,'' \emph{arXiv:2009.04332}, pp. 1--16, 2020.

\bibitem{8265165}
R.~Gray, A.~Franci, V.~Srivastava, and N.~E. Leonard, ``Multiagent
  decision-making dynamics inspired by honeybees,'' \emph{IEEE Transactions on
  Control of Network Systems}, vol.~5, no.~2, pp. 793--806, 2018.

\bibitem{9022871}
B.~Gao and L.~Pavel, ``On passivity, reinforcement learning, and higher order
  learning in multiagent finite games,'' \emph{IEEE Transactions on Automatic
  Control}, vol.~66, no.~1, pp. 121--136, 2021.

\bibitem{10.5555/3219358.3219364}
P.~Mertikopoulos and W.~H. Sandholm, ``Learning in games via reinforcement and
  regularization,'' \emph{Math. Oper. Res.}, vol.~41, no.~4, p. 1297–1324,
  2016.

\bibitem{10.2307/2285509}
M.~H. DeGroot, ``Reaching a consensus,'' \emph{Journal of the American
  Statistical Association}, vol.~69, no. 345, pp. 118--121, 1974.

\bibitem{7024142}
S.~R. Etesami and T.~Başar, ``Game-theoretic analysis of the
  {Hegselmann-Krause} model for opinion dynamics in finite dimensions,''
  \emph{IEEE Trans Automatic Control}, vol.~60, no.~7, pp. 1886--1897, 2015.

\bibitem{BAUSO2018204}
D.~Bauso and M.~Cannon, ``Consensus in opinion dynamics as a repeated game,''
  \emph{Automatica}, vol.~90, pp. 204--211, 2018.

\bibitem{Sandholm2010Population-Game}
W.~H. Sandholm, \emph{Population Games and Evolutionary Dynamics}.\hskip 1em
  plus 0.5em minus 0.4em\relax MIT Press, 2010.

\bibitem{HOFBAUER200747}
J.~Hofbauer and W.~H. Sandholm, ``Evolution in games with randomly disturbed
  payoffs,'' \emph{J Econ Theory}, vol. 132, no.~1, pp. 47--69, 2007.

\bibitem{CHEN199732}
H.-C. Chen, J.~W. Friedman, and J.-F. Thisse, ``Boundedly rational nash
  equilibrium: A probabilistic choice approach,'' \emph{Games and Economic
  Behavior}, vol.~18, no.~1, pp. 32--54, 1997.

\bibitem{10.2307/2729218}
J.~Conlisk, ``Why bounded rationality?'' \emph{Journal of Economic Literature},
  vol.~34, no.~2, pp. 669--700, 1996.

\bibitem{Franci2021}
A.~Franci, A.~Bizyaeva, S.~Park, and N.~E. Leonard, ``Analysis and control of
  agreement and disagreement opinion cascades,'' \emph{Swarm Intelligence},
  vol.~15, no.~1, pp. 47--82, 2021.

\bibitem{plaut1973derivatives}
R.~Plaut and K.~Huseyin, ``Derivatives of eigenvalues and eigenvectors in
  non-self-adjoint systems.'' \emph{AIAA Journal}, vol.~11, no.~2, pp.
  250--251, 1973.

\bibitem{golubitsky1985}
M.~Golubitsky and D.~Schaeffer, \emph{Singularities and Groups in Bifurcation
  Theory (Volume I)}.\hskip 1em plus 0.5em minus 0.4em\relax Springer-Verlag
  New York, 1985.

\end{thebibliography}

\end{document}